# PhosNetVis: A web-based tool for fast kinase-substrate enrichment analysis and interactive 2D/3D network visualizations of phosphoproteomics data

Osho Rawal[1,9], Berk Turhan[1,2,9], Irene Font Peradejordi[1,3], Shreya Chandrasekar[1,3], Selim Kalayci[1], Sacha Gnjatic[4,5], Jeffrey Johnson[6], Mehdi Bouhaddou[7,8], Zeynep H. Gümüş[1,4,10,*]

[1] Department of Genetics and Genomics, Icahn School of Medicine at Mount Sinai, New York, NY 10029, USA

[2] Faculty of Engineering and Natural Sciences, Sabanci University, Istanbul 34956, Türkiye

[3] Cornell Tech, Cornell University, New York, NY 10044, USA

[4] Marc and Jennifer Lipschultz Precision Immunology Institute, Icahn School of Medicine at Mount Sinai, New York, NY 10029, USA

[5] Department of Immunology and Immunotherapy, Icahn School of Medicine at Mount Sinai, New York, NY 10029, USA

[6] Department of Microbiology, Icahn School of Medicine at Mount Sinai, New York, NY 10029, USA

[7] Department of Microbiology, Immunology, and Molecular Genetics, University of California, Los Angeles; Los Angeles, CA 90095, USA

[8] Institute for Quantitative and Computational Biosciences, University of California, Los Angeles; Los Angeles, CA 90095, USA

[9] These authors contributed equally

[10] Lead contact

* Correspondence: zeynep.gumus@mssm.edu

**Summary**

Protein phosphorylation involves the reversible modification of a protein (substrate) residue by another protein (kinase). Liquid chromatography-mass spectrometry studies are rapidly generating massive protein phosphorylation datasets across multiple conditions. Researchers then must infer kinases responsible for changes in phosphosites of each substrate. However, tools that infer kinase-substrate interactions (KSIs) are not optimized to interactively explore the resulting large and complex networks, significant phosphosites, and states. There is thus an unmet need for a tool that facilitates user-friendly analysis, interactive exploration, visualization, and communication of phosphoproteomics datasets. We present PhosNetVis, a web-based tool for researchers of all computational skill levels to easily infer, generate and interactively explore KSI networks in 2D or 3D by streamlining phosphoproteomics data analysis steps within a single tool. PhostNetVis lowers barriers for researchers in rapidly generating high-quality visualizations to gain biological insights from their phosphoproteomics datasets. It is available at: https://gumuslab.github.io/PhosNetVis/

## Graphical Abstract

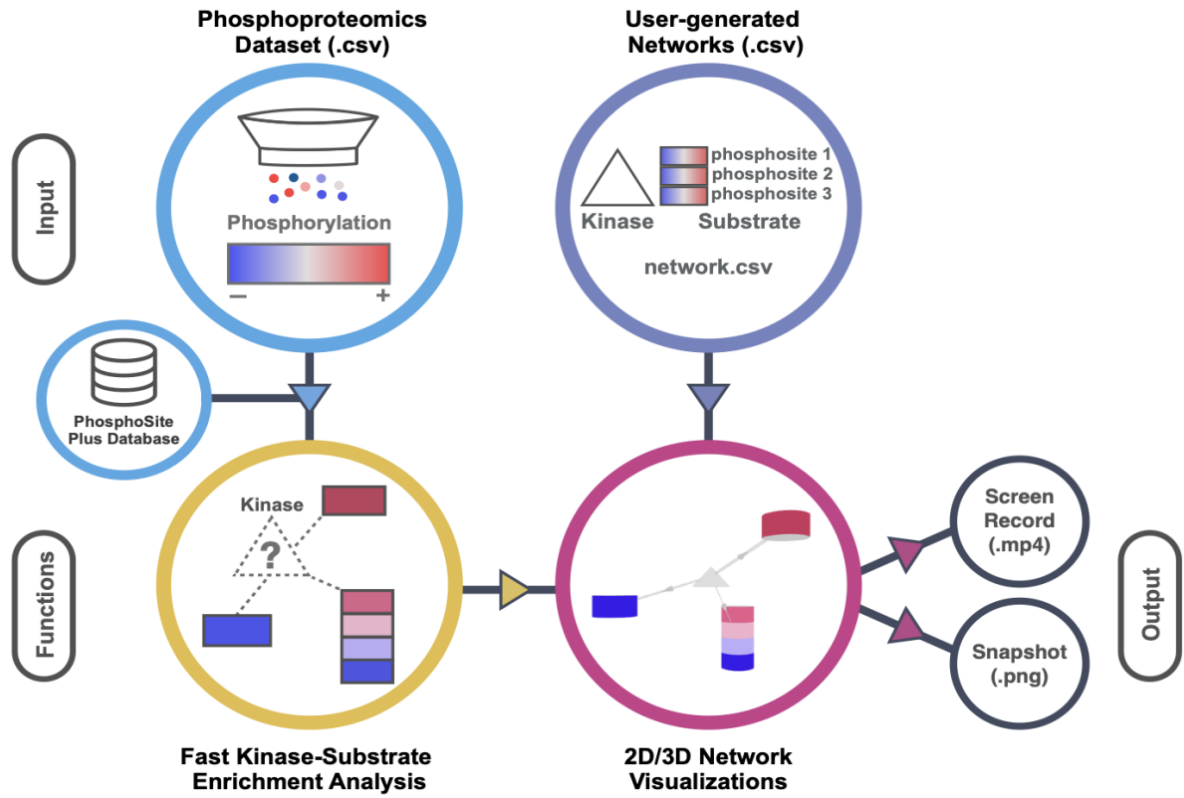

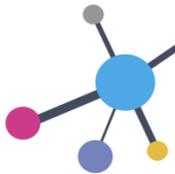

**Introduction**

Protein phosphorylation is a vital process in cellular signaling where a kinase protein modifies a residue on a substrate protein. This reversible modification can occur at multiple sites (phosphosites) on a single substrate, with different kinases targeting various substrates. Advances in liquid chromatography-mass spectrometry (LC-MS) technology have enabled the rapid generation of extensive protein phosphorylation datasets across different cellular states. To identify significant kinase-substrate interactions (KSIs) from these large datasets, bioinformatic analysis tools are essential. These tools infer which kinases are responsible for the observed changes in protein phosphorylation using available KSI database resources, and facilitate the visual exploration of the resulting KSI networks.

Over the past decade, a wide array of computational tools have been developed to infer kinases[1,2]. Many of these tools involve kinase-substrate enrichment analysis, which employs a gene-set enrichment analysis algorithm to determine if a set of query proteins is enriched in substrates known to interact with specific kinases. Commonly employed kinase enrichment tools are KEA3[3], KSEAapp[4], PhosFate Profiler[5], RoKAI[6], KSEA[7], KSEAplus[8], and pCHIPS[9]. These tools typically utilize publicly available KSI databases such as Phospho.ELM[10], PhosphoSitePlus[11], the Human Protein Reference Database (HPRD)[12], Swissprot[13], and Kinase Library[14]. PhosphoSitePlus[11] is often preferred, as it is regularly updated. As of the latest update, PhosphoSitePlus lists 10,995 KSI pairs (8,004 in human), and 23,800 phosphosites (14,663 in human) (version downloaded on Mon Oct 28 2024, from the database update of Thu Oct 17 11:21:06 EDT 2024).

After performing kinase enrichment, significant KSIs identified based on user-defined criteria are often visualized as networks, with kinases and substrates represented as nodes and their interactions as edges. Visualizing the altered phosphorylation states of phosphosites in substrate proteins is also essential. Despite the proliferation of tools to infer KSIs from phosphoproteomics datasets, currently available tools are not optimized for visualizing and generating large, complex KSI networks along with their associated phosphorylation sites and states. Furthermore, there are no dedicated tools that integrate kinase enrichment with interactive visualizations of the resulting KSI networks and phosphosites across multiple conditions.

Current network data exploration workflows involve manual adjustments to maximize usefulness, minimize clutter, and improve visualization design. This process can be improved with scripting languages; however, proteomics researchers should not need to learn programming skills to create visualizations. Even then, tools for network visualizations typically do not include the functionalities to visually represent phosphorylation data of protein residue sites and states[15]. At the same time, tools specifically designed for visual exploration of proteomics datasets do not allow users to upload and explore their own data. For instance, ProNetView[16], a web-based interactive 3D network visualization tool developed by our group, is tailored for exploring specific proteomics network data from the Clinical Proteogenomic Tumor Analysis Consortium (CPTAC)[17]. Furthermore, similar to network visualization tools, ProNetView does not visually represent the phosphorylation sites and states. One exception is the Cytoscape[18] app, Omics Visualizer[19], which allows visual representations of phosphosite data, but does not include a kinase enrichment component. At the same time, the popular web-based platform to investigate kinase activity data, PhosFate Profiler[5], only enables data exploration in tabular format, without KSI network or phosphosite data visualization functionalities. Another web-based platform, RoKAI[6], for inferring kinase activity, also does not include capabilities for enhanced network visualizations and interactive explorations. There is thus a clear need for a tool that facilitates user-friendly generation, interactive exploration, visualization, and communication of phosphoproteomics datasets.

Here, we present PhosNetVis, a freely accessible web-based tool designed for users of all computational proficiency levels to explore kinase activity from phosphoproteomics studies. PhosNetVis integrates multiple analysis steps within a single platform, allowing users to perform kinase-substrate enrichment analysis and/or build and visualize shareable interactive 2D or 3D KSI networks effortlessly. The tool provides a versatile environment, offering visual representations of protein networks, phosphorylation sites, and states of interest, along with associated differential phosphorylation and statistical significance data. Users can interact with the data through KSI network visualizations and tabular formats, all within a single interface. The KSI networks can visually include phosphorylation sites and their respective states (increased or decreased). Input is straightforward, requiring users to upload their datasets as comma-separated files on their web browser. While the PhosNetVis tool is tailored for the analysis of phosphoproteomics datasets, its adaptable architecture extends its utility to other biomolecular network applications, even in the absence of phosphorylation data. This versatility enhances its potential impact across a wide range of applications.

The PhosNetVis web portal provides detailed tutorials, an FAQ section, and interactive examples that highlight its user-friendly design and versatility. Additionally, as a resource for the research community, the portal hosts a complete catalog of KSI networks across 7 tumor immune subtypes, derived from over 1,000 tumors across 10 different cancers, sourced from the Clinical Proteomic Tumor Analysis Consortium (CPTAC) initiative[20]. As we describe in the Illustrative Example section, users can seamlessly query, visualize, interactively explore, compare, and download these KSI networks with phosphorylation state data, facilitating easy access to this extensive dataset.

**Results**

*User Research and Prototyping*
Before designing and building PhosNetVis, we conducted preliminary user research with five domain practitioners. This research involved observing their current workflows, identifying pain points, and gathering requirements for ideal features. We analyzed the user input network data and how these data were typically mapped to 2D visual elements using existing tools. To establish the general workflow and convert these mappings from 2D to 3D, we first created low-fidelity (low-fi) prototypes on paper. This approach allowed us to quickly and efficiently present our assumptions on the functionalities and workflow. Following the low-fi prototypes, we developed high-fidelity (hi-fi) prototypes using Adobe XD ([adobe.com/products/xd](adobe.com/products/xd)) and Figma ([figma.com](figma.com)). We created multiple separate prototypes for both stages, which were eventually merged into a single final design. We presented the hi-fi prototype to prospective users within the Human Immunology Project Consortium (HIPC) and Clinical Proteomic Tumor Analysis

Consortium (CPTAC). Based on their feedback, we made further improvements to both the prototype and the technical implementation.

*PhosNetVis Workflow Design*
Building KSI networks from phosphoproteomics datasets typically involves a two-step workflow: first using a kinase enrichment algorithm to establish KSI networks, and then employing visualization tools to explore and analyze these networks. PhosNetVis streamlines this process by integrating both kinase enrichment and interactive visual network exploration into a single, unified solution.

The overall architecture and main functional components of PhosNetVis are represented in Figure 1. Briefly, to infer a network of significant KSIs from a phosphoproteomics study, users input their differential phosphorylation data (log2FC) into the fast kinase substrate enrichment algorithm (fKSEA) interface of PhosNetVis. For fKSEA, PhosNetVis uses the fast gene set enrichment (FGSEA) algorithm[21], which allows having more permutations and thereby more fine-grained p-values than using standard multiple hypothesis correction methods. The fGSEA process generates a list of KSIs with their associated Benjamini & Hochberg (BH) corrected p-values, with a KSI considered significant if its BH-adjusted p-value is less than or equal to a user-defined cutoff. Users have the flexibility to adjust the input parameters and labels as needed.

The fKSEA page generates KSI network file(s) in .CSV format, which can then be explored in the network visualization page. On the network visualization page, users can visually explore the KSI network(s), examine phosphorylation sites and their states, query for specific proteins, adjust significance thresholds for differential phosphorylation, pan, zoom, and animate network changes across different states, such as multiple treatments or time points. Additionally, PhosNetVis allows for the direct upload of users' own phosphoproteomics network data, whether custom-built or from other kinase enrichment tools. This feature enables users to leverage the tool's interactive visual exploration functionalities independent of its kinase enrichment capabilities. Alternatively, users can also download the results of their fKSEA runs to visually explore the resulting network data files using other tools.

*Fast Kinase-Substrate Enrichment Analysis (fKSEA) Page*
A snapshot of the PhosNetVis fKSEA interface is shown in Figure 2 (descriptive text not shown). This page has three sections: input file format (Figure 2A), adjust parameters (Figure 2B) and upload data and run analysis (Figure 2C).

<u>*fKSEA Input File Format.*</u> Prior to using the fKSEA interface, users should process their proteomics data for differential phosphorylation analysis between two states (e.g., baseline vs. perturbed) to obtain log2 fold change (perturbed/baseline) phosphorylation (log2FC) values and associated p-values. Figure 2A illustrates the format of a sample input file, with mandatory fields marked by an asterisk (*). A sample fKSEA input file is also provided in Table S1. An input file should be in .CSV format and include at least two columns:

- Protein accession ID (identifies the protein)
- Log2FC differential phosphorylation value at a phosphosite on that protein.

Optionally, users can include additional columns:

- Unique phosphosite ID for when multiple phosphosites are involved (e.g. phosphoSitePlus ID; residue ID; or any other custom ID)
- *p*-value associated with the differential phosphorylation fold change.

*f*KSEA Input Parameters. Users can adjust the default parameters of the FGSEA algorithm[21] through the "Adjust Parameter" section (Figure 2B). These parameters include:

- Differential phosphorylation p-value cut-off (default = 0.05)
- Output BH-adjusted p-value cutoff (default=0.05)
- Minimum size of a KSI set to test (minSize) (default=15)
- Maximum size of a KSI set to test (maxSize) (default=100)
- Boundary for calculating the p-value (eps) (default=0).

For protein labels, users can choose between UniProt accession IDs or HUGO gene IDs. After uploading data and customizing parameters, the fKSEA pipeline maps the input data to KSIs within the PhosphoSitePlus database[11] [version downloaded on Mon Oct 28 2024, from the database update of Thu Oct 17 11:21:06 EDT 2024] to identify KSIs that pass user-defined significance thresholds. Additionally, each kinase is assigned an enrichment score by the FGSEA algorithm[21], which has been demonstrated in previous studies to effectively estimate kinase activities[22].

*fKSEA Output.* After fKSEA runs, the resulting KSI network is output into a .CSV file. Users are then prompted with a success message that redirects them to download the KSI network connectivity data and/or to directly visualize the network in the network visualization page. A sample fKSEA output file is provided in Table S2.

*KSI Network Visualization*
To visualize the KSI networks, users can either directly go from fKSEA page to the network visualization page, or skip the fKSEA step and input their datasets directly for network visualizations in the network data input page.

*Network Data Input Page.* This page allows users to input one or more phosphoproteomics network dataset files corresponding to different cellular states (e.g. time, treatment, exposure) for KSI network visualizations. It also includes descriptive text and a formatting guide for optional customizations of the network input files. A snapshot of this page is provided in Figure 3 (descriptive text on its website not shown).

*Network Data Input File Format.* A network input file should include at least two columns: Kinases (Kinase ID) and Substrates (Target IDs). A sample input file format is also provided in Table S3. Users can optionally include additional attributes to customize their network visualizations according to their needs. These optional attributes include:

- KinaseSize: Node size associated with each kinase.

- KinaseActivity: Kinase node color representing positive or negative direction of kinase activity.
- EdgeWeight: Custom thickness for each edge.
- EdgeHue: Edge hue changes between user-defined minimum and maximum values.
- TargetSize: Node size associated with each substrate.
- PhosphoSiteID: Unique ID of any phosphosite users deem important to visualize.
- log2FC: Phosphosite color representing the log fold change in phosphorylation at that site.
- pValue: Associated p-value of phosphorylation.

These attributes allow users to tailor their visualizations for better clarity and insight, depending on their needs. However, please note that additional customizations are not required to visually explore KSI networks generated by the fKSEA page, as these can be directly visualized within PhosNetVis. In addition, if users prefer to upload their own KSI networks, as in the CPTAC networks described in the Illustrative Examples section, the minimum requirement of PhosNetVis is a simple .CSV file that includes two columns named "KinaseID" and "TargetID". Furthermore, while the interactive network exploration page is optimized for KSI networks, by design it can accommodate visual explorations of any directed biomolecular network, as long as edge directionality for the source and target nodes is provided by using the "KinaseID" and "Target ID" column names.

*KSI Network Visualization Interface.* Once the KSI network files are generated in the fKSEA page, or uploaded directly through the network data input page, the user is directed to the interactive network visualization interface. PhosNetVis interactive interface is in user-optional 2D or 3D. For a comparative understanding of the 2D versus 3D layouts, Figure 4 provides two snapshots of the same KSI network generated with PhosNetVis: Panel A is a rendering in 3D layout, and Panel B shows it in 2D layout. The 2D view also includes node labels. This KSI network visualization interface features three components:

*Control Panel* (Figure 4, Panels A/B, left corner). This interactive panel allows users to seamlessly transition between 2D and 3D network layouts; query for specific nodes; dynamically adjust node color thresholds based on phosphorylation levels and fold-change criteria; customize the background color; toggle the visibility of phosphorylation site partitions and node labels; enter or exit fullscreen mode; capture screenshots; and restore the network visualization to default settings. Additionally, users can perform gene enrichment analyses by sending their list of genes in the network to the Enrichr tool[23–25] with a simple click, enabling more in-depth biological insights.

*Visualized Network* (Figure 4, Panels A/B middle). Users can interact with the 2D or 3D network by rotating, zooming in/out; dragging nodes to different positions or selecting nodes or edges. Substrates are represented as cylinders. The height of each cylinder depends on the number of phosphosites harbored by the substrate it represents, with each differentially phosphorylated phosphosite depicted as a slice of the cylinder. The color of each cylinder slice indicates its level of differential phosphorylation, ranging from blue (reduced phosphorylation) to red (increased phosphorylation). A maximum 10 of the most differentially altered phosphosites are shown per substrate. Kinases are represented as triangles in the 2D configuration, and as cones in the 3D configuration. For kinases, if they harbor differentially phosphorylated phosphosites, these are represented as slices below the gray triangles in the 2D configuration, colored according to the differential phosphorylation levels. In the 3D configuration, kinases without differentially phosphorylated phosphosites are shown as gray cones. If they do have such phosphosites, the cones contain slices colored according to the differential phosphorylation levels.

*Pop-up node details table* (Figure 4 Panel A bottom left corner). Double-clicking on a node either in 2D or 3D opens an interactive table displaying detailed information on the node. This includes the node type (kinase or substrate), and for each differentially phosphorylated site on that node, its position, differential phosphorylation fold change value (perturbation/baseline), and differential phosphorylation p-value. Additionally, it includes hyperlinks to the PhosphoSitePlus database[11] for more detailed protein information.

*Legend* (Figure 4, Panels A/B, bottom right corner). This section explains the user-selected range of Phosphorylation Log Fold Change values. Each cylinder slice representing a phosphosite is colored based on the maximum negative and positive differential phosphorylation levels in the study. Lighter hues indicate smaller values, while darker hues correspond to larger values, with blue representing the most negative and red the most positive differentially phosphorylated phosphosite. If optional parameters were included in the input file, the legend also shows the range for these attributes, such as Node Size, Edge Color, and Edge Weight (edge thickness). In the given example (Figure 4, Panels A/B), the input file includes only fold change information, so the legend displays only the range of differential fold change of phosphorylation.

This setup ensures that users can thoroughly explore and analyze their KSI networks with a high degree of interactivity and customization.

<u>*Network Visualization Output File Format.*</u> Users can take snapshots of their network visualizations, or screen-record animations of differential phosphorylation changes across multiple states (from multiple input files). Snapshots are saved in .PNG file format. Screen recordings are saved in .MP4 file format.

*Illustrative Examples*
We illustrate the functionalities of PhosNetVis KSI network visualization interface with two case studies. Both are available for interactive exploration within the tool webportal. The portal also includes several toy datasets customized to help users get familiar with different KSI network attributes. By exploring these examples, users can visually interact with the networks, understand tool features and experiment with further customizations of their datasets.

<u>*SARS-CoV-2 KSI network.*</u> This example is curated from a study that examined the global phosphorylation landscape of SARS-CoV-2 infection[26]. In this study, Vero-E6 cells, an African green monkey cell line, were infected with SARS-CoV-2, the virus responsible for COVID-19. Cells were harvested at various time points post-infection (2, 4, 8, 12, and 24 hours). Phosphoproteomics analysis using liquid chromatography-mass spectrometry (LC-MS) quantified 4,624 human-orthologous phosphorylation sites across 3,036 human-orthologous proteins. A screenshot of this KSI network at the 24 hour time point is shown in Figure 4, with Panel A in 3D and Panel B in 2D.

One significant finding in this study was the increased activity of Casein Kinase II (*CSNK2A1*) following infection. This was evidenced by a marked increase in the abundance of known *CSNK2A1* phosphorylation sites post-infection. *CSNK2A1* was localized within filopodia protrusions that emerged from the cell surface during infection, suggesting a role in filopodia formation. Further analysis of phosphorylation sites affecting cytoskeletal changes revealed increased phosphorylation of *CTNNA1, HDAC2, HMGA1, HMGN1*, and *STAT1* proteins at sites known to be associated with cytoskeleton remodeling. Specifically, phosphorylation was observed at *CTNNA1:S641, HDAC2:S394, HMGA1:S102-103, HMGN1:S7, STAT1:S727*). Zooming in on *CSNK2A1* in this network and inspecting its substrates and known phosphosites visually confirms this finding. These interactions were identified by running the fKSEA algorithm on the phosphoproteomics data, and are depicted in the magnified subnetwork of *CSNK2A1* (Figure 4, Panels A/B, zoomed-in circles). Clicking on one of its substrates, *HMGA1*, brings up a pop-up table of its differentially phosphorylated phosphosites, S102 and S103, along with their log2 fold change (infection/mock) and p-values (Figure 4, Panel A, bottom left). For further details on *HMGA1*, the table links to PhosphoSitePlus[11].

This KSI network is available for interactive exploration in PhosNetVis at: https://gumuslab.github.io/PhosNetVis/existing-networks.html. Users can inspect the phosphorylation status of the network at a time point of interest by simply using the dropdown menu in the Control Panel (Figure 4, Panels A/B, top left), toggle across different time points, or inspect an animation of the changes in phosphorylation within the KSI network over time by selecting the *Animate & Record* tab from the Control Panel. These functionalities help users pinpoint potential mechanisms of interest for further studies, providing a dynamic and detailed view of the KSI network and the phosphorylation events during SARS-CoV-2 infection.

*Clinical Proteomic Tumor Analysis Consortium (CPTAC) pan-cancer immune subtype KSI networks.* Recently, through the Clinical Proteomic Tumor Analysis Consortium (CPTAC), we investigated the immune landscape of over 1,000 tumors from ten different cancer types[20]. This effort aimed to enhance the understanding of immune cell surveillance mechanisms and the various strategies tumors use to evade immune responses. Using CPTAC's comprehensive pan-cancer proteogenomic data[27], this study identified and characterized seven unique immune subtypes, and by analyzing kinase activities within these subtypes, it uncovered potential therapeutic targets specific to each subtype. Here, we provide the full catalog of PhosNetVis interactive visualizations of KSI networks of each immune subtype both as a resource for the research community and as use case examples. This PhosNetVis application skipped the fKSEA Page. Instead, the networks were derived from simply mapping the list of proteins (both kinases and substrates) in each immune subtype to the PhosphoSitePlus database[11] to derive their connecting edges between the kinases and the phosphosites in the CPTAC immune study[20]. Then, the adjacency matrices of all immune subtype KSI networks were directly uploaded into PhosNetVis using its Network Data Input Page to generate their interactive visualizations.

The PhosNetVis catalog of these networks allows easy queries, interactive visualizations, exploration, and download, and is available through the PhosNetVis Existing Networks link at: https://gumuslab.github.io/PhosNetVis/cptac-vis.html. Using PhosNetVis, users can on-the-fly comparatively analyze KSIs and differential phosphorylations across these 7 immune subtypes. For example, Figure 5 illustrates two of the largest-connected network snapshots in a 2D layout.

*Panel A* shows the most active pan-cancer immune subtype (CD8+/IFNG+), which leads to the upregulation of PRKACA and the phosphorylation of its substrates. *Panel B* depicts the least active cluster (CD8-/IFNG-), resulting in the upregulation of cell-cycle kinases such as CDK1 and the phosphorylation of its substrates.

These CPTAC networks provide good case studies on how users can interact with the networks in 2D or 3D layout, and then directly embed their network snapshots in 2D into publications, as we illustrate in Figure 5. The drag function allows users to easily modify the network layout, emphasizing specific parts of the network as needed. In Figure 5, Panels A and B, quick manual alterations in the network connectivities and layouts highlight the main kinases more clearly. These visualizations enable researchers to pinpoint potential mechanisms of interest for follow-up studies. Furthermore, the catalog of these pan-cancer immune subtype KSI networks enables the broader community of researchers to explore complex KSI network datasets from the CPTAC initiative.

**Discussion**

Large-scale phosphoproteomics experiments are characterizing protein phosphorylation sites and states across conditions to better understand cellular signaling in health and disease. While numerous tools are dedicated to infer kinases from these datasets, there is a need for integrated visualization tools that allow users to explore the resulting KSI networks across multiple conditions. Here, we introduced PhosNetVis, an interactive web-based tool that enables users to perform fast Kinase-Substrate Enrichment Analysis (fKSEA), and then to visually explore, interpret and communicate the inferred KSI networks and their associated phosphorylation data across multiple states. PhosNetVis provides a private analysis environment, as all operations during KSI network visualization take place exclusively within the user's local browser, without relying on an external server or third-party site. In the case of fKSEA, datasets undergo processing on the protected RStudio Connect server at Mount Sinai (https://rstudio-connect.hpc.mssm.edu) to guarantee data privacy. By integrating fKSEA and network visualization in a single platform, PhosNetVis simplifies the workflow for analyzing phosphoproteomics data, making it accessible to users of varying computational proficiency levels.

PhosNetVis features an easy-to-navigate graphical user interface for KSI networks that visually represents phosphorylation sites and their respective states across multiple experiments. Users can explore KSI network visualizations, switch between different network states (i.e. time points or treatments), toggle between 2D or 3D visualizations, and adjust network parameters. The tool allows users to query nodes of interest for detailed phosphosite information, change differential phosphorylation parameters, rotate, pan, zoom in/out, drag nodes, download the network snapshots or record animations of network changes over time or conditions. While PhosNetVis is specifically designed for phosphoproteomics analysis, its adaptable architecture extends its utility to various biomolecular network applications, even when phosphorylation data are unavailable. This adaptability broadens its potential impact across diverse research problems, enabling researchers to visualize and interpret complex biological data in a dynamic and interactive manner. One limitation is that while fKSEA analysis employs the most recent version of the most popularly used kinase-substrate database, PhosPhoSitePlus[11], the upstream kinases of a majority of phosphosites are still unknown. Furthermore, 20% of those kinases that are known are in the upstream of the majority of known phosphosites [28]. This, however, is a limitation within the field, and is an active area of research, with studies ongoing to reveal the full extent of human KSIs.

Overall, PhosNetVis lowers the barriers between complex phosphoproteomics data and researchers who want rapid, intuitive, and high-quality tools to infer kinase activity and thereby visually explore kinase-substrate interaction networks at multiple phosphorylation sites and states. This will empower investigators in translating rich datasets into biological insights and clinical applications. PhosNetVis is freely accessible at https://gumuslab.github.io/PhosNetVis. Its website hosts detailed tutorials, an FAQ page and a variety of use case examples, including the full catalog of different KSI networks within 7 different tumor immune subtypes derived from a pan-cancer analysis of more than 1,000 CPTAC tumors across 10 different cancers.

**Methods**

*Website Implementation*
*fKSEA Page.* We developed a user-friendly interface for fKSEA using HTML, CSS, and Bootstrap 4. FGSEA is performed using a Plumber API (https://www.rplumber.io/), deployed on the RStudio Connect server at Mount Sinai (https://rstudio-connect.hpc.mssm.edu). Users upload their CSV files and initiate the analysis via a simple interface, which sends an API request to the server to perform FGSEA. This setup ensures efficient and seamless processing of kinase enrichment analysis, making it accessible and convenient for researchers.Once fKSEA API call is finished, users are directed to download the network data files or to directly visualize the files..

*Network Data Input Page.* Users can upload one or more network data files through this page. Network data files can be the outputs from PhosNetVis fKSEA page analysis or custom-generated. The page provides detailed guidelines for network data input formats for custom-generated files. For CSV file parsing and data processing, we integrated the PapaParse library (https://www.papaparse.com/) to enable users to concurrently upload and process multiple CSV files. Once network data CSV files are uploaded, PhosNetVis transforms each input .CSV file into the JavaScript Object Notation (JSON) format using the Danfo.JS package (danfo.jsdata.org/).

*Interactive Network Visualization Page.* Once network data files are uploaded, all networks are visualized in the network visualization page. We built this interface mainly with HTML and CSS. We used JavaScript for the Document Object Model (DOM) element manipulations, and the Bootstrap library to implement responsive layouts and to customize page architectures (getbootstrap.com). To create a screen-resizable responsive canvas where the 3D network visualization is rendered, we integrated the element-resize-detector library (github.com/wnr/element-resize-detector)[30].

To prioritize user interactions and control, we included several graphical user interface (GUI) elements to the network visualization interface (Figure 4, Panels A and B). Briefly, to enable users to interactively modify their visualizations, adjust parameters and conduct node queries, we implemented a GUI control panel by using the Tweakpane.js library

([cocopon.github.io/tweakpane/](cocopon.github.io/tweakpane/)) (Upper left corner in Figure 4, Panels A and B). To enhance user interaction through specific node queries, we used the Fstdropdown.js library ([github.com/VirtusX/fstdropdown](github.com/VirtusX/fstdropdown)). To offer users a 3D network visualization experience (Figure 4, Panel A), we used the 3d-force-graph library ([github.com/vasturiano/3d-force-graph](github.com/vasturiano/3d-force-graph)) to render the network layouts in 3D. The library leverages the WebGL-based Three.js ([threejs.org](threejs.org)) library to create interactive force-directed graphs in 3D space. Users can easily customize node colors and labels in 3D, for which we used RainbowVis-JS ([github.com/anomal/RainbowVis-JS](github.com/anomal/RainbowVis-JS)) and three-spritetext ([github.com/vasturiano/three-spritetext](github.com/vasturiano/three-spritetext)) libraries, respectively. In addition to 3D, PhosNetVis also offers a 2D network visualization option (Figure 4, Panel B). To develop the 2D network visualization option, we implemented the force-graph library ([github.com/vasturiano/force-graph](github.com/vasturiano/force-graph)), which uses a force-directed layout algorithm to present network layouts effectively on an HTML5 canvas. Finally, for users to export visualizations, we implemented jsScreenRecorder ([github.com/manan657/jsScreenRecorder](github.com/manan657/jsScreenRecorder)), a custom JavaScript script designed to capture and record screen interactions, particularly beneficial after rendering the networks in 3D. By implementing these tools and libraries, PhosNetVis ensures a comprehensive and user-friendly network visualization experience, supporting researchers in their exploration of KSI networks and associated phosphorylation data.

*Tutorials, Interactive Examples and Toy Datasets*
To guide users, the documentation platform of PhosNetVis ([https://gumuslab.github.io/phosnetvis-docs/](https://gumuslab.github.io/phosnetvis-docs/)) features tutorials that cover fKSEA and network visualization, as well as instructions on input data formatting. In addition, we provide several interactive examples and toy datasets at [github.com/GumusLab/PhosNetVis-DataExamples](github.com/GumusLab/PhosNetVis-DataExamples).

**Resource availability**

*Lead contact*
Dr. Zeynep H. Gümüş can be reached by email ([zeynep.gumus@mssm.edu](zeynep.gumus@mssm.edu)).

*Materials availability*
PhosNetVis is hosted on GitHub Pages and is freely available with no login requirements at [https://gumuslab.github.io/PhosNetVis](https://gumuslab.github.io/PhosNetVis). For additional materials, please contact the lead contact.

*Data and code availability*
PhosNetVis source code is available on GitHub repository ([github.com/GumusLab/PhosNetVis](github.com/GumusLab/PhosNetVis)) at [https://doi.org/10.5281/zenodo.14215570](https://doi.org/10.5281/zenodo.14215570)[29], and is released under the GNU Affero General Public License AGPL-3.0 ([https://www.gnu.org/licenses/agpl-3.0.en.html](https://www.gnu.org/licenses/agpl-3.0.en.html)), and is also available under a commercial license for enterprises seeking additional features or avoiding AGPL obligations.

**Acknowledgments**

ZHG gratefully acknowledges support from NIH R33 CA263705-01; SG from NIH U24 CA224319, U01 DK124165, and P01 CA196521; and MB from NIH K99 AI163868. Authors gratefully acknowledge valuable feedback from investigators within NIH funded Human

Immunology Project Consortium (HIPC) and Clinical Proteomic Tumor Analysis Consortium (CPTAC), with special thanks to Francesca Petralia for kindly sharing the CPTAC pan-cancer immune subtype networks with the study team.

**Author contributions**

O.R: writing - reviewing & editing, visualization, software, investigation. B.T: writing - original draft, writing - reviewing & editing, visualization, software, investigation. I.F.P: writing - original draft, visualization, software, investigation. S.C: writing - original draft, visualization, software, investigation. S.G: writing - reviewing, visualization. S.K: visualization, software. J. J: writing - original draft, writing - reviewing & editing, visualization, software, conceptualization. M.B.L: writing- reviewing, visualization. Z.H.G: writing - original draft, writing - reviewing & editing, conceptualization, visualization, supervision.

**Declaration of interests**

SG reports other research funding from Boehringer-Ingelheim, Bristol-Myers Squibb, Celgene, Genentech, Regeneron, and Takeda, and consulting from Taiho Pharmaceuticals, not related to this study. All other authors declare no competing interests.

**Figure titles and legends:**

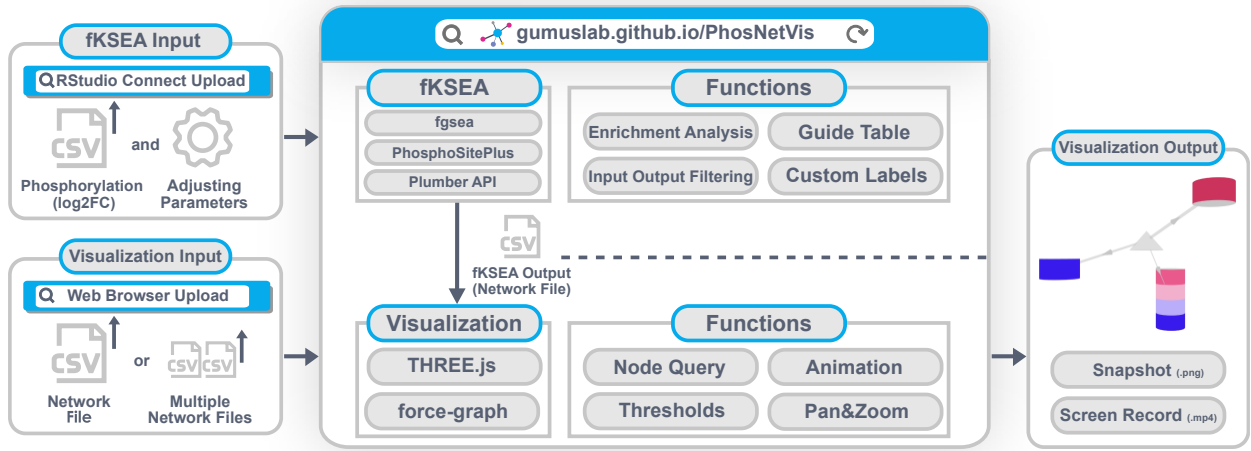

**Figure 1. PhosNetVis overall architecture and main components.**

**Figure 2. fKSEA page snapshot.** A.) Input File Format section describes input .CSV file format for fKSEA input. B.) Adjust Parameters section enables users to adjust parameters based on their needs before running the FGSEA algorithm. C.) Upload Data & Run analysis section enables users to upload their input file from their local directory; run fKSEA; download the generated KSI network file and visualize the generated network.

**FORMAT AND UPLOAD YOUR DATA**

| KinaseID (Required) | TargetID (Required) | KinaseSize (Optional) | KinaseActivity (Optional) | EdgeWeight (Optional) | EdgeHue (Optional) | TargetSize (Optional) | PhosphoSiteID (Optional)* | log2FC (Optional) (Phosphorylation) | pValue (Optional) (Phosphorylation) |
|---|---|---|---|---|---|---|---|---|---|
| CDK1 | NUCKS1 | 1.2 | 0.16 | 1 | 1.44 | 0.9 | S19 | -0.1 | 0.05 |
| CDK1 | NUCKS1 | 1.2 | 0.16 | 1 | 1.44 | 0.9 | T202 | -4.32 | 0.006 |
| CDK1 | CDK1 | 1.2 | 0.16 | 1.5 | 4.28 | 1.2 | T161 | 1.47 | 0.12 |
| CDK1 | NPM1 | 1.2 | 0.16 | 3.8 | 0.07 | 1.12 | T78 | 2 | 0.0235 |
| CSNK2A1 | TOP2A | 4.56 | -2.27 | 1.67 | 2 | 2.98 | S29 | -3.059 | 0.03 |
| MAPK3 | TOP2A | 1 | 1.2 | 2.28 | 1.5 | 2.98 | K431 | -2.51 | 0.01 |

*PhosphositePlusID or Residue ID

If you have more than one dataset, select the number of datasets and make sure to upload all of them!

Number of Datasets: 1    Choose Files No file chosen    Upload Sample Data

**UPLOAD**

or

**FAST KINASE-SUBSTRATE ENRICHMENT ANALYSIS**    TUTORIAL

**Figure 3. Snapshot of the Network Visualization Input Page** (descriptive text not shown). This page allows users to upload network files for visualization. Users can upload their custom network files or customize those generated by the fKSEA page. For comparative analyses across networks, users can upload multiple files by selecting the number of datasets they want to upload. This page also displays a table that provides the required network data format. To visualize network data, users need a .CSV file with at least two columns: 1) Kinases (KinaseID) and 2) Target Nodes (TargetID) for substrates. Optionally, users can add additional columns to customize the node and edge attributes.

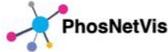
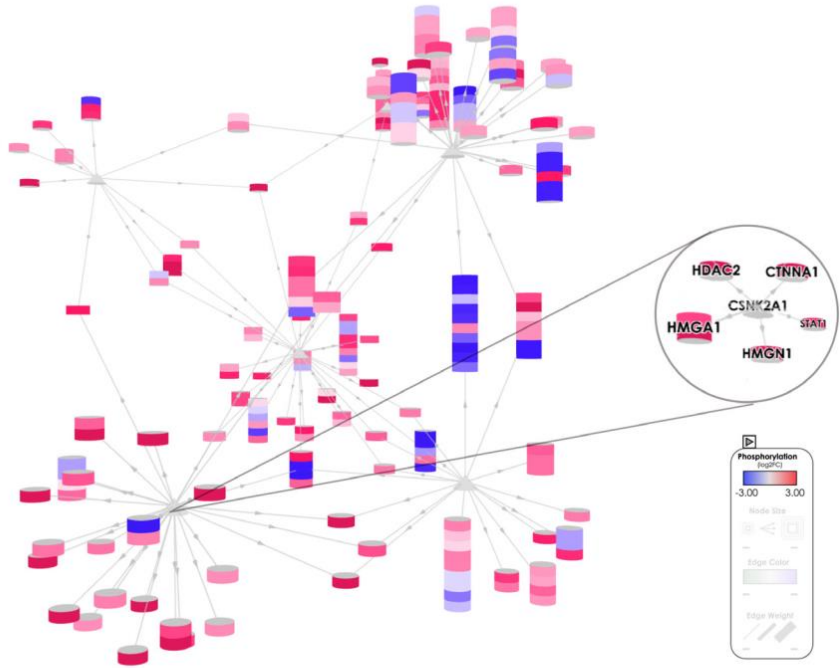
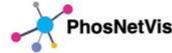
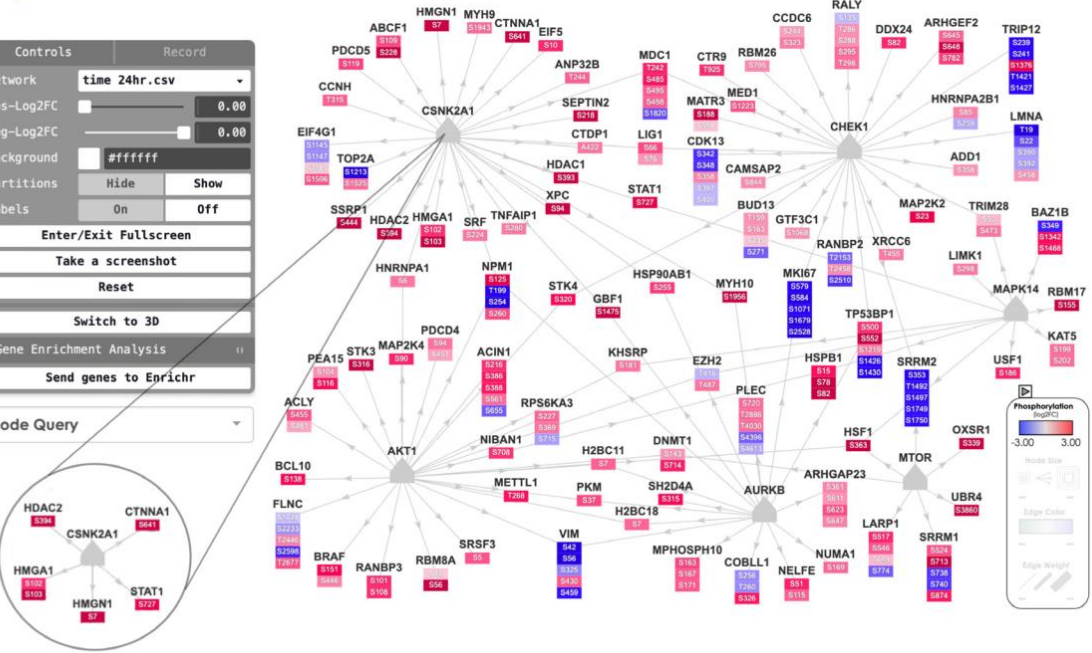

**Figure 4. Snapshots of network visualization interface featuring the phosphorylation landscape of SARS-CoV-2 infection.** This page allows users to view and interact with the network curated from the Global Phosphorylation Landscape of SARS-CoV-2 Infection (Bouhaddou et al., 2020). Users can rotate the network, zoom in/out, pan through, see a node information by double clicking on it, or reset the view by double clicking on the background. Additionally, it provides options for the user to switch between 3D or 2D network. Users can

also send their genes list to the Enrichr tool[23–25] for gene enrichment analyses, for further exploration of enriched pathways and functional gene groups. Double-clicking on any node opens an interactive table that displays detailed information on the node as shown in Panel A bottom left for node *HMGA1*. **A)** Snapshot of the 3D view of the network, including a magnified view of a CSNK2A1-central subnetwork in 3D. This subnetwork highlights phosphorylation sites known to be associated with cytoskeleton remodeling, filtered from the main network. **B)** Snapshot of the 2D view of the same network. In each panel, the left corner shows the control panel; the middle displays the network; the bottom right corner contains the legend. The starred section indicates the CSNK2A1 phosphorylation subnetwork, including a magnified view of the same CSNK2A1 node with select interactors in a subnetwork in 2D. These features enable comprehensive exploration and analysis of the phosphorylation events during SARS-CoV-2 infection.

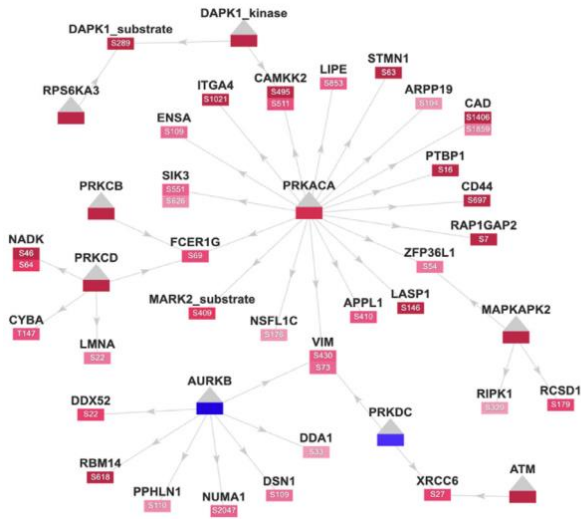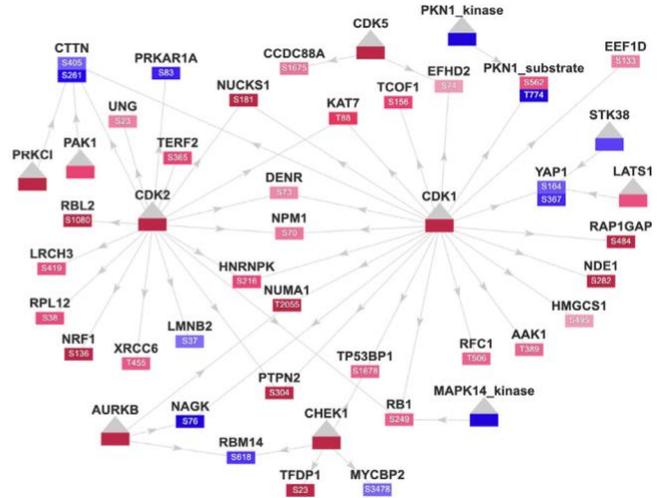

**Figure 5. Snapshots of interactive 2D network visualizations for CPTAC pan-cancer immune subtypes.** Kinases, and their specific altered phosphosites, their phosphorylation levels and positions in each altered substrate are clearly shown **A) CD8+/IFNG+ network.** Visualization clearly depicts upregulation in the global abundance of the kinase PRKACA (center) and the phospho-abundance of its substrates in red; **B) CD8-/IFNG- subnetwork.** The global proteomic expression of cell-cycle kinases such as CDK1 and the phospho-abundance of its substrates are upregulated in red.

## SUPPLEMENTAL TABLES

**Table S1.** Fast Kinase-Substrate Enrichment Analysis (fKSEA) Input File Format. [(R): Required, (O): Optional]

| ProteinAccession (R) | log2FC (R) | PhosphoSiteID (O) | pValue (O) |
|---|---|---|---|
| P02545 | -1.8 | S22 | 0.0025 |
| P02545 | -0.95 | S390 | 0.03 |
| P17612 | 2.1 | T198 | 0.028 |
| P08670 | -3.48 | S56 | 0.024 |

**Table S2.** Fast Kinase-Substrate Enrichment Analysis (fKSEA) Output File Format. [(H|A): HUGO or Accession ID, (IE): If Exists]

| KinaseID (H|A) | TargetID (H|A) | PhosphoSiteID (IE) | log2FC | KinaseActivity | pValue |
|---|---|---|---|---|---|
| AKT1 | LMNA | S22 | -1.8 | -1.3 | 0.0025 |
| AKT1 | LMNA | S390 | -0.95 | -1.3 | 0.03 |
| PRKACA | PRKACA | T198 | 2.1 | 0.7 | 0.028 |
| PRKACA | VIM | S56 | -3.48 | 0.7 | 0.024 |

**Table S3.** Visualization Input File Format. [(R): Required, (O): Optional, (P): Phosphorylation]

| Kinase ID (R) | Target ID (R) | Kinase Size (O) | Kinase Activity (O) | Edge Weight (O) | Edge Hue (O) | Target Size (O) | PhosphoSite ID (O) | log2FC (O) (P) | pValue (O) (P) |
|---|---|---|---|---|---|---|---|---|---|
| AKT1 | LMNA | 1.2 | -1.3 | 1 | 1.44 | 0.9 | S22 | -1.8 | 0.0025 |
| AKT1 | LMNA | 1.2 | -1.3 | 1 | 1.44 | 0.9 | S390 | -0.95 | 0.03 |
| PRKACA | PRKACA | 3 | 0.7 | 1.5 | 4.28 | 3 | T198 | 2.1 | 0.028 |
| PRKACA | VIM | 3 | 0.7 | 3.8 | 0.07 | 1.12 | S56 | -3.48 | 0.024 |